\documentstyle[preprint,aps]{revtex}
\begin{document}
\draft
\title{\bf GENERALISED RAYCHAUDHURI EQUATIONS FOR STRINGS 
IN THE PRESENCE OF AN ANTISYMMETRIC TENSOR FIELD}
\author{Sayan Kar \thanks{Electronic Address :
sayan@iopb.ernet.in} \footnote{Address after Sept. 1, 1996 :
IUCAA, Post Bag 4, Ganeshkhind, Pune 411 007, INDIA}}
\address{Institute of Physics\\
Sachivalaya Marg, Bhubaneswar--751005, INDIA}
\maketitle
\parshape=1 0.75in 5.5in
\begin{abstract}
The generalised Raychaudhuri equations derived by Capovilla and
Guven are exclusively for extremal, timelike Nambu--Goto membranes.
In this article, we construct the corresponding equations for
string world--sheets in the presence of a background Kalb--Ramond field.
We analyse the full set of equations by concentrating on special cases
in which the generalised shear
or the generalised rotation or both are set to zero. 
If only the generalised shear is set to zero then it is possible 
to identify the
components of the generalised rotation with the projections of the field
strength of the Kalb--Ramond potential. 
\end{abstract}
\parshape=1 0.75in 5.5in
PACS number(s) : 11.27. +d, 11.10.Kk


\section{Introduction}
   
The importance of the consequences resulting out of the analysis of the
Raychaudhuri equations for timelike/null geodesic congruences {\cite
{akr:prd55}}, {\cite{wald:85}}, {\cite{he:cup72}} in the 
proof of the singularity theorems of General Relativity (GR)
{\cite{he:cup72}}, {\cite{wald:85}}
is a well known fact today. At the same time, string and membrane
theories have been extremely popular among theoretical physicists of
different specialisations. Apart from it being a candidate for 
quantum gravity and unification of forces there are a multitude
of situations ranging from particle physics to biology where
membranes or strings play a pivotal role in describing
a system or analysing a phenomenon.   
Given the diverse areas in which strings and membranes 
are applied it is perhaps worthwhile to know the generalisations
of the basic equations for curves (such as the geodesic equation,
the Jacobi equation and the Raychaudhuri equation) to the case of
surfaces.
Earlier papers by Guven{\cite{guv:prd93}}, Larsen and Frolov
{\cite{lf:npb94} and Carter{\cite{cart:prd94} have succeeded
in generalising the equation of geodesic deviation to the
case of timelike surfaces.
More recently, Capovilla and Guven {\cite{cg:prd95}}
have generalised the Raychaudhuri equations for timelike
geodesic congruences to the case of families of $D$ dimensional, timelike
, extremal, Nambu--Goto surfaces
in an $N$ dimensional background. Subsequently, several illustrative
examples of these equations were constructed by this author in
{\cite{sk:prd95,sk:prd96}}. The generalisation of the notion of
geodesic focussing to families of surfaces was  also introduced and
discussed in some detail.

However, it is necessary to realise that there are several other 
actions that arise in string and membrane theories which are
different from the Nambu--Goto/Polyakov actions. Differences
arise in various ways--the presence of background fields other
than gravity {\cite{kr:prd}}, supersymmetrisations {\cite{sup:gsw}}
, rigidity corrections {\cite{rig:zac}} and so on.
What are the generalisations of the Raychaudhuri equations for 
these actions ? In this paper we attempt to analyse one such
action--the one in the presence of a background antisymmetric 
tensor field (the Kalb--Ramond field).

The generalised Raychaudhuri equations now contain many extra
new terms which embody several nontrivialities. 
We discuss special cases in which either the shear or the
rotation or both are set to zero and thereby obtain
simplified sets of equations. For the case in which only the
shear is set to zero it is possible to
identify the projections of the antisymmetric tensor field
with the components of the rotation in a special way.

The paper is organised as follows. 
Section II provides some background material based on the 
paper by Capovilla and Guven. In the third section we write
down and analyse the equations for the case of strings in the
presence of an antisymmetric tensor field. Finally, in Section IV
we conclude with remarks on future directions.
 
\section{Background}

In this section we briefly review the work of Capovilla and Guven
{\cite{cg:prd95}} on the generalisation of the Raychaudhuri equations.

The geometric objects under consideration are $D$ dimensional, timelike
surfaces embedded in an $N$ dimensional background. We denote $x^{\mu}$
($\mu = 0,1,2,......N-1$)  as the background spacetime coordinates and 
$\xi^{a}$ ($a = 0,1,2,......D-1$) as the surface coordinates. 
Tangents and normals are defined in the usual way and we can construct
an orthonormal, spacetime basis $\{E^{\mu}_{a}, n^{\mu}_{i}\}$ at each point
on the surface. The projections of the spacetime covariant derivatives of
$E^{\mu}_{a}$ and $n^{\mu}_{i}$ along the surface tangents define the
extrinsic curvatures, Ricci rotation coefficients and the twist potentials
through the standard Gauss--Weingarten formulae given as : 
 
\begin{eqnarray}
D_{a}E_{b} = {\gamma}_{ab}^{c}E_{c} -K _{ab}^{i}n_{i} \\
D_{a}n_{i} = K_{ab}^{i}E^{b} + {\omega}^{ij}_{a}n_{j}
\end{eqnarray}

where $D_{a} \equiv E^{\mu}_{a} D_{\mu}$ ( $D_{\mu}$ being the usual 
spacetime covariant derivative). The quantities $K_{ab}^{i}$ (extrinsic 
curvature), ${\omega}_{a}^{ij}$ and
 ${\gamma}_{ab}^{c}$ are defined as :

\begin{eqnarray}
K_{ab}^{i} = -g(D_{a}E_{b}, n^{i}) = K_{ba}^{i} \\
{\omega}_{a}^{ij} = g(D_{a}n^{i},n^{j})\\
{\gamma}_{abc} = g(D_{a}E_{b}, E_{c}) = -{\gamma}_{acb}
\end{eqnarray}

In order to analyse deformations normal to the worldsheet we need to
consider the normal gradients of the spacetime basis set.
The corresponding analogs of the Gauss--Weingarten
equations are :

\begin{eqnarray}
D_{i}E_{a} = J_{aij}n^{j} + S_{abi}E^{b} \\
D_{i}n_{j} = -J_{aij}E^{a} + {\gamma}^{k}_{ij}n_{k}
\end{eqnarray}

where $D_{i} \equiv n^{\mu}_{i} D_{\mu}$ ( $D_{\mu}$.
The quantities $J_{a}^{ij}$, $S_{abi}$ and
 ${\gamma}^{k}_{ij}$ are defined as :

\begin{eqnarray}
S^{i}_{ab} = g(D^{i}E_{a}, E_{b}) = -S^{i}_{ba}\\ 
{\gamma}_{ijk} = g(D_{i}n_{j}, n_{k}) = -{\gamma}_{ikj}\\
J_{a}^{ij} = g(D^{i}E_{a},n^{j}) 
\end{eqnarray}

The quantity $J_{aij}$ is the most crucial object in our discussion
because it is, in a sense, a measure of the allowed orthogonal deformations
of the world--surface which is necessary in obtaining information of the
behaviour of families of surfaces. One can think of $J_{aij}$ as a quantity
which essentially is a  
surface analog of the gradient of the tangent vector field in the case 
of geodesics. The major point of difference with the case of geodesics
is the fact that one has one quantity along each of the tangential
directions on the surface.
 
The evolution equations for the $J_{aij}$ are the generalised Raychaudhuri
equations.

Instead of taking the proper time derivative of $J_{aij}$
we look at the covariant worldsheet derivative of this quantity. This turns 
out to be (for details see Appendix of {\cite{cg:prd95}}) given as :

\begin{equation}
{\tilde{\nabla}}_{b}J^{ij}_{a} = -{\tilde{\nabla}}^{i}K_{ab}^{j}
-{J_{b}^{i}}_{k}{J_{a}^{kj}} - K_{bc}^{i}K^{cj}_{a} - g(R(E_{b},n^{i})E_{a},
n^{j})
\end{equation}

where the extrinsic curvature tensor components are $K_{ab}^{i} =
-g_{\mu\nu}E^{\alpha}_{a}(D_{\alpha}E^{\mu}_{b})n^{\nu i}$

On tracing over worldsheet indices we get
\begin{equation}
{\tilde{\nabla}}_{a}J^{aij} = -{J_{a}^{i}}_{k}{J^{akj}} - K_{ac}^{i}K^{acj}
 - g(R(E_{a},n^{i})E^{a},
n^{j})
\end{equation}

where we have used the equation for extremal, timelike, Nambu--Goto
membranes (i.e. $K^{i} = 0$). As we shall see in the next section,
it is this condition which will change and thereby introduce all the
nontrivialities when we introduce a background antisymmetric tensor
field.

The antisymmetric part of (12) is given as :

\begin{equation}
{\tilde{\nabla}}_{b}J^{ij}_{a}- {\tilde{\nabla}}_{a}J^{ij}_{b}
= G_{ab}^{ij}
\end{equation}

where $g(R(Y_{1},Y_{2})Y_{3},Y_{4}) = R_{\alpha\beta\mu\nu}Y_{1}^{\alpha}
Y_{2}^{\beta}Y_{3}^{\mu}Y_{4}^{\nu}$ and

\begin{equation}
G_{ab}^{ij} = -{J_{b}^{i}}_{k}{J_{a}^{kj}} - K_{bc}^{i}K^{cj}_{a} - g(R(E_{b},n^{i})E_{a},
n^{j}) - (a\rightarrow b)
\end{equation}

One can further split $J_{aij}$ into its symmetric traceless, trace and
antisymmetric parts ($J_{a}^{ij} = {\Sigma}_{a}^{ij} + {\Lambda}_{a}^{ij}
+ \frac{1}{N-D}{\delta}^{ij}{\theta_{a}}$) and obtain the evolution equations
for each of these quantities. The one we shall be concerned with mostly
is given as 

\begin{equation}
\Delta \gamma + {\frac{1}{2}}{\partial}_{a}\gamma{\partial}^{a}\gamma + 
(M^{2})^{i}_{i} = 0
\end{equation} 

with the quantity $(M^{2})^{ij}$ given as :

\begin{equation}
(M^{2})^{ij} = K_{ab}^{i}K^{abj} +
R_{\mu\nu\rho\sigma}E^{\mu}_{a}n^{\nu i}E^{\rho a}n^{\sigma j}
\end{equation} 

 ${\nabla}_{a}$ is the worldsheet covariant derivative ($\Delta = {\nabla^{a}
 \nabla_{a}}$) and ${\partial}_{a}\gamma={\theta}_{a}$. 
  Notice that 
 we have set ${\Sigma}_{a}^{ij}$ and 
${\Lambda}^{ij}_{a}$ equal to zero. This is possible only if the symmetric 
traceless part of $(M^{2})^{ij}$ is zero. One can check this by looking 
at the full set of generalized Raychaudhuri equations involving 
$\Sigma^{ij}_{a}$, ${\Lambda}^{ij}_{a}$ and $\theta_{a}$ [4].
For geodesic curves the usual Raychaudhuri
equations can be obtained by noting that $K^{i}_{00} = 0$, the $J_{aij}$ are 
related to their spacetime counterparts $J_{\mu\nu a}$ through
the relation $J_{\mu\nu a} = n^{i}_{\mu}n^{j}_{\nu}J_{aij}$,
and the $\theta$ is defined by contracting with the projection 
tensor $h_{\mu\nu}$.

The $\theta _{a}$ or $\gamma$ basically tell us how the
spacetime basis vectors change along the normal directions  as we
move along the surface. If ${\theta}_{a}$ diverges somewhere
, it induces a divergence in $J_{aij}$ , which, in turn means
that the gradients of the spacetime basis along the normals have
a discontinuity. Thus the family of worldsheets meet along a
curve and a cusp/kink is formed. This, we claim, is a  
focussing effect for extremal surfaces analogous to 
geodesic focussing in GR where families of geodesics focus at a point 
if certain specific conditions on the matter stress energy are
obeyed. 

\section{Strings in the presence of an antisymmetric tensor field}

In the presence of a background anti--symmetric tensor field 
(the Kalb--Ramond field) the usual Nambu--Goto/Polyakov action 
gets generalised by the addition of an extra term.
We therefore have :

\begin{equation}
 S = S_{NG} + \int {\epsilon}^{ca}B_{\mu\nu}{\partial}_{c}x^{\mu}{\partial}
_{a}x^{\nu} d^{2}{\xi}
\end{equation}

The field equation resulting from the variation of the action w.r.t. 
$x^{\mu}$ is given as :

\begin{equation}
K^{i} = \frac{1}{2} n^{i}_{\lambda}g^{\alpha\lambda}{\epsilon}^{ca}{\partial}
_{c}x^{\beta}{\partial}_{a}x^{\nu} H_{\beta\alpha\nu}
\end{equation}

where 

\begin{equation}
H_{\mu\nu\alpha} = \partial_{\mu}B_{\nu\alpha} + \partial_{\nu}B_{\alpha\mu}
+ \partial_{\alpha}B_{\mu\nu}
\end{equation}

We can also write the field equation in the following alternative form
by choosing the conformal gauge. In that case we have

\begin{equation}
\partial_{a}\partial^{a} x^{\lambda} + \Gamma^{\lambda}_{\rho\sigma}
\partial_{a}x^
{\rho} \partial^{a}x^{\sigma} = -\frac{1}{2}\epsilon^{ca}\partial_{c}x^{\beta}
\partial_{a}x^{\nu}g^{\alpha\lambda}H_{\alpha\nu\beta}
\end{equation}

The constraints, ofcourse, remain the same as for the Nambu--Goto action
for extremal surfaces.

Given the above field equation we can ask -- When does the worldsheet 
retain its minimal character even though background fields/potentials
may be present ? This implies investigating the conditions under which
the R.H.S of the equation becomes zero. 

Firstly, if $H_{\mu\nu\rho} = 0$ i.e. $B_{\mu\nu}$ is pure gauge the R. H. S. 
disappears and we are left with an equation which is the same as the 
original Nambu--Goto equation.

Additionally, if we choose stationary strings then it is not necessary
that all components of $H_{\mu\nu\rho}$ be zero. Given the stationary
string ansatz ($t=\tau, x^{i} = x^{i}({\sigma})$ ) we find that 

\begin{equation}
K^{i} = n^{qi}x^{p\prime} H_{0qp}
\end{equation}

Thus, if $H_{0qp} = 0 \forall p,q$ then $K_{i} = 0$ and the worldsheet
remains a minimal surface. 

On the other hand, for circular strings in a spherically
symmetric background ($t=t(\tau), l = l(\tau), \theta =
\frac{\pi}{2}, \phi =\sigma$) we have 

\begin{equation}
K^{i} = -n^{\alpha i}\dot t H_{\alpha 03} - n^{\alpha i} \dot l H_{\alpha 13}
\end{equation}

Therefore, with

\begin{equation}
H_{103} = H_{203} = H_{213} = 0
\end{equation}

we end up with $K_{i} = 0$ and the minimality of the surface is retained.
 
We now need to evaluate the quantity ${\tilde {\nabla}}^{i}K^{j}$. It is this
term which will add up to the generalised Raychaudhuri equation for the
usual Nambu--Goto action 
and result in the major differences. 

After some amount of algebra one finds the following expression for 
${\tilde {\nabla}}^{i}K^{j}$.

\begin{eqnarray}
{\tilde {\nabla}}^{i}K^{j} = \frac{1}{2}\epsilon^{ca} J^{ij}_{m}
\left ( E^{m}_{\lambda}E^{\beta}_{c}E^{\nu}_{a}H^{\lambda}_{\beta\nu}\right )
 \nonumber \\
 - \epsilon^{ca}S_{cd}^{i} \left ( E^{d\beta}E^{\nu}_{a}n^{j}_{\lambda}
H^{\lambda}_{\beta\nu} \right ) - \epsilon^{ca}J_{c}^{ik}
\left ( n^{j}_{\lambda}n^{\beta}_{k}
E^{\nu}_{a} H^{\lambda}_{\beta\nu} \right )
\end{eqnarray}

 The quantities in brackets in the above expression can be thought of as 
projections of $H_{\mu\nu\rho}$ along normal and tangential directions.
Thus, we end up with

\begin{equation}
{\tilde {\nabla}}^{i}K^{j} = \epsilon^{ca} \left ( \frac{1}{2} J^{ij}_{m}
H^{m}_{ca} - S_{cd}^{i} H^{jd}_{a} - J_{c}^{ik}H^{j}_{ka} \right )
\end{equation}

where we have assumed 

\begin{equation}
D^{i}H_{\mu\nu\rho} = 0
\end{equation}

Note that the first term can be very easily shown to be equal to 
zero since the worldsheet indices take only two values ($\sigma, \tau$)
. The second term for the case of the string can be written as follows :

\begin{equation}
S_{cd}^{i}\epsilon^{ca}H^{jd}_{a} = \epsilon^{\tau\sigma}S_{\tau\sigma}^{i}
H^{j\sigma}_{\sigma} + \epsilon^{\sigma\tau}S_{\sigma\tau}^{i}
H^{j\tau}_{\tau}
\end{equation}

and it is obvious that it is equal to zero by the antisymmetry of
$H^{ia}_{b}$ w.r.t. $a,b$ indices.
Hence only the last term in (25) survives. 

We are now, in a position to input this into the term containing
${\tilde {\nabla}}^{i}K^{j}$ in the generalised Raychaudhuri equation
for string theory in the presence of a background antisymmetric tensor field.

The equation obtained by tracing the expression for 
$\tilde{\nabla_{b} }J_{a}^{ij}$ with 
respect to the worldsheet indices is given as :

\begin{equation} 
{\tilde{\nabla}}_{a}J^{aij} = -{J_{a}^{i}}_{k}{J^{akj}} - K_{ac}^{i}K^{acj}
 +  \epsilon_{ca} J_{c}^{ik}H^{j}_{ka} - g(R(E_{a},n^{i})E^{a},
n^{j})
\end{equation}

We now use the splitting of $J_{a}^{ij}$ into its symmetric traceless
, antisymmetric and trace parts :

\begin{equation}
 J_{a}^{ij} = \Sigma_{a}^{ij}
+ \Lambda_{a}^{ij} + \frac{1}{N-D}\delta^{ij} \theta_{a}
\end{equation}

The equations for each of these quantities -- $\theta_{a}$,
$\Sigma_{aij}$ and $\Lambda_{aij}$ turn out to be :

\begin{equation}
\tilde{\nabla}_{a}\theta^{a} + {\Sigma_{a}}^{i}_{k}\Sigma^{aki} 
+ {\Lambda_{a}}^{i}_{k}\Lambda^{aki} + \frac{1}{N-D}\theta_{a}\theta^{a}
+ (M^{2})^{i}_{i} - \epsilon^{ca}\Lambda_{c}^{ik}H_{ika} = 0
\end{equation}

\begin{eqnarray}
\tilde{\nabla}_{a}\Lambda^{aij} - \Sigma^{a[i}_{k}\Sigma_{a}^{j]k} 
+ \Lambda^{ak[i} {\Lambda_{a}^{j]}}_{k} - 2{\Lambda^{ak[i}}{\Sigma_{a}^{j]}}
_{k}
- \epsilon^{ca}\Sigma_{c}^{k[i}H^{j]}_{ka} \nonumber \\
+ \epsilon^{ca} \Lambda_{c}^{k[i}H^{j]}_{ka} + \frac{2}{N-D}\epsilon^{ca}
\theta_{c}H^{ij}_{a} + \frac{2}{N-D}\theta^{a}\Lambda_{a}^{ij} = 0
\end{eqnarray}

\begin{eqnarray}
\tilde{\nabla}_{a}\Sigma^{aij} + \left ( \Lambda^{aik}\Lambda_{ak}^{j}
+ \Sigma^{aik}\Sigma_{ak}^{j} \right ) ^{str} 
+ \frac{2}{N-D} \Sigma_{a}^{ij}\theta^{a} \nonumber \\
 + \left [ (M^{2})^{ij}\right ] ^{str}
- \epsilon^{ca}\Sigma_{c}^{k(i}H^{j)}_{ka} + \epsilon^{ca}\Lambda_{c}^{k(i}
H^{j)}_{ka} = 0
\end{eqnarray}
 
The first fact to note in the above equations is that if we set $\Lambda$
and $\Sigma$ both equal to zero the second equation leads to a relation 
between
the $\theta_{a}$ and the $H^{ij}_{a}$ which is given as :

\begin{equation}
\epsilon^{ca}
\theta_{c}H^{ij}_{a} = 0
\end{equation}

This constrains the choice of $\theta_{a}$ and $H^{ij}_{a}$. If 
$H_{\sigma}^{ij} = H_{\tau}^{ij}$ then, as a consequence we
have $\theta_{\sigma} = \theta_{\tau}$. Therefore, the $\theta_{a}\theta^{a}$
term in the first equation vanishes and we end up with the
following equation :

\begin{equation}
\tilde{\nabla}_{a}\theta^{a} + (M^{2})^{i}_{i} = 0
\end{equation}

Furthermore, if we assume only $\Lambda_{a}^{ij} = 0$ and 

\begin{equation}
\Sigma_{\sigma}^{ij} = \Sigma_{\tau}^{ij}\quad ; \quad 
H_{\sigma}^{ij} = H_{\tau}^{ij}
\end{equation}

we end up with Eqn (33) and a similar equation for $\Sigma_{aij}$
:
\begin{equation}
\tilde{\nabla}_{a}\Sigma^{aij} - \left [ (M^{2})^{ij}\right]^{s.tr}
= 0
\end{equation}
 
Finally, we set only $\Sigma$ equal to zero. Therefore we end up
with the following algebraic relation resulting from the third
equation.

\begin{equation}
\left ( \Lambda^{aik}\Lambda_{ak}^{j}\right )^{str} + \epsilon^{ca}\Lambda
_{c}^{k(i}H^{j)}_{ka} = 0
\end{equation}

assuming that the symmetric traceless part of the object $(M^{2})^{ij}$
is equal to zero. 

This above algebraic relation can be satisfied (for the string case)
if we assume the following to hold true :

\begin{equation}
{\Lambda_{\tau^{j}}}_{k} = H^{j}_{k\sigma}\quad ; \quad \Lambda^{j}_{\sigma k}
= H^{j}_{k\tau}
\end{equation}

Thus, a geometric quantity $\Lambda_{a}^{ij}$ has been related to a
physical object-- the field strength of the Kalb--Ramond potential. 
It would be worthwhile trying to understand how the generalised 
rotation has its physical meaning in the projections of the
$H_{\mu\nu\rho}$.
 
With this identification of the $\Lambda$ with the projections of
$H$ we find that the second equation reduces to the simple equation

\begin{equation}
\tilde{\nabla}_{a}\Lambda^{a}_{ij} = 0
\end{equation}

and the equation for the generalised expansion $\theta$ turns out to be 

\begin{equation}
\tilde{\nabla}_{a}\theta^{a} + \frac{1}{N-D}\theta_{a}\theta^{a}
+ (M^{2})^{i}_{i} = 0
\end{equation}

Let us now turn to the antisymmetric equations. Recall that the $\tilde
{\nabla}^{i}K^{j}$ term does not contribute to these equations. They are
given by :

\begin{equation}
2\tilde{\nabla}_{[a}\Lambda_{b]}^{ij} = -2\Lambda_{[a}^{k[i}
{\Lambda_{b]}^{j]}}_{k} - 2{\Sigma_{[a}^{k[i}}
{\Sigma_{b]}^{j]}}_{k} -\Omega_{ab}^{ij}
\end{equation}

\begin{equation}
2\partial_{[a}\theta_{b]} = 0
\end{equation}

\begin{equation}
2\tilde{\nabla}_{[a}\Sigma_{b]}^{ij} = -2\left ( \Lambda_{a[}^{ik}\Lambda
_{b]k}^{j} + \Sigma_{[a}^{ik}\Sigma_{b]k}^{j} \right )^{str}
+ 4\Lambda_{[a}^{k(i}\Lambda_{b]k}^{j)}
\end{equation}

First, let us look at the case with $\Sigma_{\sigma}^{ij} =
\Sigma_{\tau}^{ij}$, $H_{\sigma}^{ij} = H_{\tau}^{ij}$ and 
$\Lambda_{a}^{ij} = 0$. With these assumptions, we can easily check
that the antisymmetric equations essentially reduce to one
equation given by :

\begin{equation}
\tilde{\nabla}_{a}\Sigma_{aij} = 0
\end{equation}

However, one of the traced equations (36) matches with the above equation
only if $((M^{2})^{ij})^{s.tr} = 0$. 

For the case in which $\Sigma_{a}^{ij} = 0$,
the identification of the $H$ with the $\Lambda$ results in some constraints
on the $H$ or the $\Lambda$ which have to be satisfied in order to have
a consistent solution of the full set of equations.
These constraints turn out to be as follows.
 
The second of these equations results in the relation $\theta_{a} = \partial
_{a}\gamma$. The third is satisfied identically with the assumption of 
the relation between $H$ and $\Lambda$.
The first equation (with the choice $\Omega_{a}^{ij} = 0$)
yields an extra constraint on the $H$ which reads as
follows :

\begin{equation}
\tilde{\nabla}_{[a} \Lambda_{b]}^{ij} = 2H_{[b}^{kj} H_{a]}^{i}{k}
\end{equation}

For the string world--sheet in a background four dimensional spacetime 
one ends up with the following constraints on $H$.

\begin{eqnarray}
\tilde{\nabla}_{\sigma}H^{ij}_{\tau} = \tilde{\nabla}_{\tau}H^{ij}_{\sigma}\\
\tilde{\nabla}_{\sigma}H^{ij}_{\sigma} = \tilde{\nabla}_{\tau}H^{ij}_{\tau}
\end{eqnarray}

This is because there are only two normals to the world--sheet and 
the $i,j$ indices in $H^{ij}_{a}$ are antisymmetric. The first of these 
two equations follows from (39) while the second one is a descendant of
(44).

If we further assume that the string--worldsheet is flat then the 
covariantised world--sheet derivatives are reduced to ordinary
derivatives on the world--sheet. The pair of first order partial differential
equations can therefore be thought of as a single second order wave
equation for either $H^{ij}_{\sigma}$ or $H^{ij}_{\tau}$ given as :

\begin{equation}
\frac{{\partial}^{2}H^{ij}_{\tau,\sigma}}{{\partial}\tau^{2}}
- \frac{{\partial}^{2}H^{ij}_{\tau,\sigma}}{{\partial}\sigma^{2}}
 = 0
\end{equation}

Therefore, with the identification of the projections of the $H_{\mu\nu\rho}$
field with the generalised rotation $\Lambda_{aij}$ we end up with an equation
for the $\gamma$ and two constraints on the $H$ field. This is all we need to
solve in order to analyse focussing effects for worldsheets which are
extremal solutions of the Nambu--Goto action with an antisymmetric tensor
field added to it. Infact, apart from the extrinsic curvature which will
now contain some objects related to the $H$ field and the constraints on the
$H$ we have nothing more to analyse except the usual generalised Raychaudhuri
equation for the $\theta$. The crucial result of this paper is the 
demonstration of the fact that the generalised rotation can indded be related 
to the projections of the $H_{\mu\nu\rho}$ field. This in turn leads us
to a system of equations which are tractable.

One might ask -- what are the generalised Raychaudhuri equations for
strings in a background three dimensional spacetime. Note that in this
case the string worldsheet is a hypersurface and therefore considerable
simplifications occur. One can very easily show that $\tilde{\nabla}^{i}
K^{j}$ is identically zero (the term containing a product of
normals vanishes because of the fact that there is only one 
normal now and $H^{ij}_{a}$ is antisymmetric in its 
$i,j$ indices. Therefore the generalised
Raychaudhuri equations are the same as for the case without an
antisymmetric tensor field. 

\section{Concluding Remarks}

The aim of this paper has been to derive the generalised Raychaudhuri
equations for strings in the presence of a background antisymmetric
tensor field. We have analysed several special cases by
choosing the shear, the rotation or both to zero. It turns out that if
the shear is set to zero then it
is possible to identify the
projections of the field strength of the Kalb--Ramond potential
with the generalised rotation. This, infact demonstrates that
a geometric object can be related to a physical quantity
. Recall, that in the geodesic case one could give a physical
meaning to the rotation. In the paper of Capovilla and Guven
such a physical meaning for the generalised shear or rotation
was lacking. We have, in this paper, been able to make some progress 
along this
direction for atleast one of these quantities.

A multitude of open issues remain in this area. As an extension
to this paper one can work out the generalised Raychaudhuri
equations for other actions which were mentioned in the
Introduction. Apart from this, one has to understand the
issue of focussing of surfaces in a better and more general way
without referring to specific examples. Thereafter, one can address
the question of spacetime singularities and their relation
to string focussing effects. 

In drawing parallels with the basic equations of GR--geodesic
equation, deviation equation, Raychaudhuri equations
and the Einstein equation one now notices that in the context
of strings we actually have the first three. The fourth one
is however not there. However, recall that GR as a theory has
the unique feature-- the geodesic equation (trajectory of test particles)
can be derived from the Einstein equations. Therefore, one can frame 
the question--what is the 'Einstein equation' which will lead to
the string equations of motion? 
An answer to this question will perhaps help us to understand the
relation between strings, gravity and spacetime geometry in a 
novel way.
 

\end{document}